\documentclass[sigconf]{acmart}
\usepackage{siunitx}
\usepackage{mathtools}
\usepackage{enumitem}

\AtBeginDocument{%
  }

\setcopyright{acmlicensed}
\copyrightyear{2026}
\acmYear{2026}
\setcopyright{cc}
\setcctype{by}
\acmConference[MUM '26]{25th International Conference on Mobile and Ubiquitous Multimedia}{December 01--04, 2026}{Glasgow, United Kingdom}
\acmBooktitle{25th International Conference on Mobile and Ubiquitous Multimedia (MUM '26), December 01--04, 2026, Glasgow, United Kingdom}
\acmDOI{10.1145/3861188.3861241}
\acmISBN{979-8-4007-2791-7/2026/12}

\acmConference[MUM '26]{International Conference on Mobile and Ubiquitous Multimedia}{December 01--04,2026}{Glasgow, United Kingdom}

\begin{document}

\title[Sight-Restriction-Based Collision Prevention in VR]{Immersive Sight-Restriction-Based Collision Prevention Using Generic Particle Systems in Virtual Reality}

\author{Annalisa Degenhard}
\email{annalisa.degenhard@uni-ulm.de}
\orcid{0000-0002-3190-0496}
\affiliation{%
  \institution{Ulm University}
  \city{Ulm}
  \country{Germany}
}

\author{Annika Stampf}
\email{annika.stampf@uni-ulm.de}
\orcid{0000-0001-5539-4957}
\affiliation{%
  \institution{Ulm University}
  \city{Ulm}
  \country{Germany}
}

\author{Michael Rietzler}
\email{michael.rietzler@uni-ulm.de}
\orcid{0000-0003-2599-8308}
\affiliation{%
  \institution{Ulm University}
  \city{Ulm}
  \country{Germany}
}

\author{Enrico Rukzio}
\email{enrico.rukzio@uni-ulm.de}
\orcid{0000-0002-4213-2226}
\affiliation{%
  \institution{Ulm University}
  \city{Ulm}
  \country{Germany}
}

\renewcommand{\shortauthors}{Degenhard et al.}


\begin{abstract}
    Collision prevention systems (CPS) help VR users navigate their physical surroundings safely. Most approaches overlay conceptually distinct grids or add solid virtual counterparts of physical obstacles; both can reveal the physical space, as they either stand out visually or behave inconsistently during locomotion. We explore sight restriction as an alternative: a standard particle system visualizes the play area boundary as a fog-like cloud whose intensity scales with proximity and viewing direction. A pre-study ($N=10$) determined the distance at which the visualization must appear. In a main study ($N=11$), we compared this particle-based CPS with a grid-based solution, focusing on locomotion behavior: across $253$ approach episodes, users interrupted their approach after a comparable distance once either visualization appeared. Although the particle boundary was visible three times as often, participants reported higher presence. We release our implementation as open source to support further work on immersive collision prevention.
\end{abstract}




\begin{teaserfigure}
 \centering
  \includegraphics[width=0.9\textwidth]{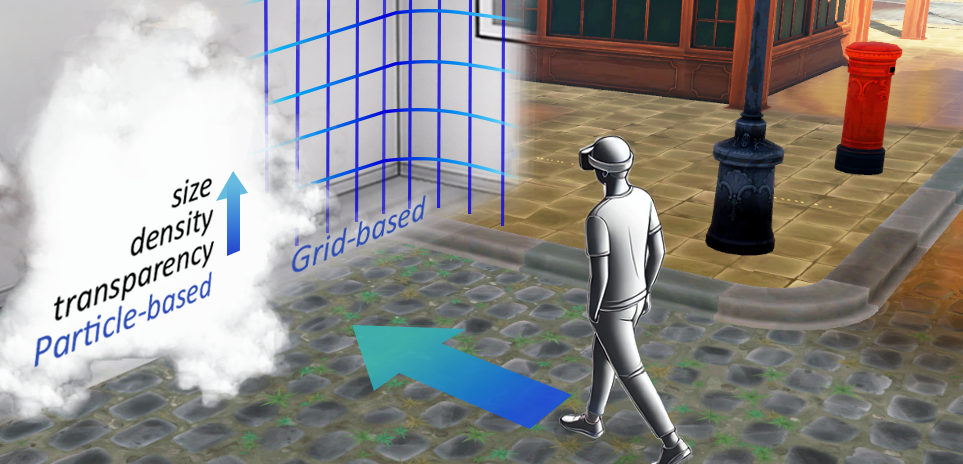}
  \caption{We explore the potential of sight restriction to keep virtual reality users inside their physical space without breaking presence. A standard particle system visualizes the play area boundary as a fog-like cloud that intensifies with proximity. Across two user studies, users stopped just as reliably as with grid-based visualizations and although the particle boundary was visible three times as often, they reported higher presence.}
  \Description{Example of our particle-based collision prevention system on the left, and a standard grid-based solution on the right. The background shows the virtual environment used within the two presented user studies.}
  \label{fig:teaser}
\end{teaserfigure}

\maketitle

\section{\MakeUppercase{Introduction}}
Collision prevention systems (CPSs), such as Meta's Guardian~\cite{meta_set_2024} or Valve's Chaperone ~\cite{chaperone_steam_nodate}, are essential for safe virtual reality (VR) use~\cite{huang_improving_2018}. They inform users about upcoming boundaries of their physical environment and prompt them to reorient. While it appears to be easy to effectively communicate these prompts, for instance, through a switch to passthrough, this would be at the cost of the immersive experience. In fact, CPSs often use virtual boundary visualizations in the first place to avoid reminding users of their physical surroundings. However, to communicate the prompt to reorient effectively yet immersive CPSs have to provide the flexibility to adapt to diverse scenarios, with respect to the virtual environment (VE), physical environment, and users' current movement. Locomotion can cause different changes in the virtual and physical environment, which make it challenging to provide a logically coherent CPS visualization that dynamically adapts accordingly. Logical inconsistencies between CPS visualization and virtual environment, however, may break plausibility and presence, respectively~\cite{slater_virtual_2000}. The perfect tradeoff between presence and effectiveness has not yet been found. State-of-the-art built-in systems mostly use grid-based boundary visualizations that appear at a certain proximity to the boundary~\cite{meta_set_2024,chaperone_steam_nodate}. Their standardized appearance may be beneficial for efficient communication of the boundaries' location, as confusion with visualizations of the actual VE is unlikely. However, this clear distinction will also likely cause breaks in presence. Another common approach is to trigger reorientation through virtual counterparts of physical obstacles~\cite{sra_procedurally_2016}. Building on intuitive collision avoidance, this approach also works with contextualized visualizations. However, it has limitations for teleportation-based VR applications, as this implies the sudden appearance of the virtual counterparts at the target location.
Another prevention approach is based on sight restrictions, such as reducing the field of view (FOV)~\cite{wu_combining_2019}. Restricting sight is conceptually a promising approach, as in reality, people also intuitively reduce their pace with increasing sight restrictions~\cite{jeon_evacuation_2011,yan_effects_2014}. However, to our knowledge, no prior work has investigated a contextualized sight-restriction-based CPS. 
In reality, smoke and fog are two prominent causes of temporarily reduced sight. The higher their density, the less people can orient themselves, to the point where navigation is barely possible. The diverse and dynamic visual appearance of so-called aerosols also offers great advantages over solid virtual counterparts. We are used to different types of aerosols as well as different ways of their formation~\cite{hinds2022aerosol}. Aerosols can be integrated into a wide range of virtual environments without introducing rigid geometric structures that are visually inconsistent with the scene. 
%Aerosols can appear and disappear in the air for no visually apparent reason, and so may they disappear again. 
Their shape and appearance depend on the context and can change dynamically due to gusts of wind or the like. This provides a set of attributes that are needed for boundary visualizations.
The question remains, however, whether aerosols can be used in VR to effectively communicate upcoming boundaries of physical space and prevent collisions while still providing an immersive alternative to standard grid-based solutions. Avoiding aerosols is not necessarily natural, as most aerosols can be passed through without harm in reality. Davis et al. showed that increasing smoke density in VR induces slowing the user's speed~\cite{davis_investigating_2023}. However, the task was still to traverse the aerosol. How users would intuitively respond to sight restrictions has not yet been investigated, especially at more severe levels. Rather, it was found that for CPS in general, users may intentionally cross boundary visualizations, because they have some sort of mental model of the physical environment~\cite{tseng_understanding_2024}. As this finding applied to grid-based solutions that resemble the shape of the configured play area, it is unclear whether users would still apply this behavior when using more neutral cloud-based CPSs. We hence sought to answer the following research questions: 

\begin{itemize}
    \item \textbf{RQ1} How should a particle-based CPS be designed to be reliably perceived and to communicate the boundary's location accurately? (Pre-study, $N=10$) 
    \item \textbf{RQ2} To what extent do a particle-based CPS and a grid-based CPS differ in user safety, measured through locomotion behavior when approaching the boundary, boundary crossings, and reported collision anxiety? (Main study, $N=11$)
    \item \textbf{RQ3} To what extent can particle-based CPS increase presence compared to a grid-based CPS? (Main study, $N=11$)
\end{itemize}

\noindent For an easily implementable and customizable solution, we designed a particle-based CPS (PCPS) using a simple particle system from Unity. For customizability, we focused the design optimization on general aspects such as the timing and manner of formation, as well as the appearance depending on the user's distance from the actual obstacle. The particles' appearance was kept flexible and can be chosen according to the individual use case. Using a simple collection task, we compared our design solution with the grid-based CPS (GCPS) of Meta~\cite{meta_set_2024}. Our work provides the following contributions~\cite{wobbrockResearchContributionsHumancomputer2016}: 

\begin{itemize}
    \item A pre-study ($N=10$) to determine user requirements for PCPS.
    \item A user study ($N=11$) that compares a GCPS with our PCPS design considering safety and presence.
    \item The Unity implementation\footnote{\url{https://github.com/annalisaDegenhard/ParticleBasedCollisionPrevention.git}}, including the parameterized particle system, the proximity- and gaze-dependent visibility model, and the collision logging components - released as open source to enable replication and reuse.
\end{itemize}

\section{\MakeUppercase{Related work}}
%Underlying problem
When navigating the VE through movements in the physical world while visually disconnected from it, disorientation in the real world is almost inevitable. CPSs in VR prevent users from collisions with physical obstacles or boundaries ~\cite{cucher_virtual_2023,jerald2015vrbook,meta_set_2024}. 
%Challenges
The challenge for CPSs is to provide safety while interfering with the actual VR experience as little as possible to avoid a reduction of presence. Arguably, the simplest solution to ensure safety would be to visualize physical surroundings in VR permanently. However, this would not only lead to a higher cognitive load~\cite{JOKINEN2020102376} but would also compromise the plausibility of the VE, as locomotion interfaces may lead to different movement translations in VR compared to the physical world~\cite{di_luca_locomotion_2021}. Plausibility, in this case, is defined as the illusion that what is apparently happening is really happening~\cite{slater_place_2009}. A common solution is an urgency-sensitive adaptation of the visualization based on the user's proximity to it~\cite{huang2018depth}. Starting with a non-intrusive small visualization, the salience of the visual stimulus is increased as the user approaches the boundary. 
This, in turn, requires balanced timing and effective high-urgency communication to ensure timely redirection by the user~\cite{jelonek2023vrtoer,cirio_walking_2012}. Many state-of-the-art solutions switch to pass-through when the user gets too close to a boundary ~\cite{meta_passthrough_2024}. The drawback is that users are reminded of being confronted with a simulation rather than a real environment.
%Presence
Presence, the subjective experience of being in one place even when physically situated in another~\cite{witmer1994measuring}, is considered one of the most important qualities of VR~\cite{greengard2019virtual} and develops when both place illusion and plausibility are given~\cite{slater_place_2009}. Being reminded of the physical environment may therefore be effective, but also causes a break in presence~\cite{slater_virtual_2000,garau_temporal_2008}, which is hard to achieve again once lost~\cite{slater_place_2009}. CPSs should thus minimize irritation while still ensuring player safety, which is challenging, as previous research has shown that most incidents in VR are caused by presence (e.g.~\cite{jelonek2023vrtoer}, referred to as immersion by the authors).

So far, there are two types of collision prevention strategies: implicit and explicit prevention. 
%Implicit prevention -> limited applicability in everyday life
Implicit prevention tries to prevent any approach to physical boundaries by e.g. redirected walking~\cite{rietzler_rethinking_2018,razzaque2002redirected,sun2018infinite,williams_exploring_2007,bachmann_collision_2013}. However, the average domestic physical area used for VR, often called play area, is smaller and includes much more hindering furniture than typical laboratory settings where redirected walking has been tested~\cite{jelonek2023vrtoer}. 
%Explicit prevention
An alternative CPS approach is based on explicit prevention by integrating obstacle-related information into the VE~\cite{jerald2015vrbook,sra_procedurally_2016,chung_vrcat_2022,valentini_improving_2020,afonso2011aural}. Most explicit CPS are based on the integration of visualizations that communicate the desired obstacle information~\cite{sra_procedurally_2016,chung_vrcat_2022}. Other approaches are feedback through other sensory modalities, like haptics or audio \cite{afonso2011aural,boldt1028intrusive,valkov2019vibrotactile}, proximity-sensitive restrictions~\cite{wu_combining_2019}, or the bottom-up generation of VEs based on scans of the physical environment~\cite{cheng_vroamer_2019}. 

\noindent Previous work has already shown an improvement in presence using contextualized CPSs compared to the GCPS of Vive~\cite{valentini_improving_2020}. The contextualized solutions resulted in greater involvement, spatial presence, and realism than the GCPS. However, since the proposed solution was based on the integration of virtual counterparts, the findings are probably limited to VR solutions that solely use real walking for locomotion. If users are able to, for example, teleport, the appearance of their virtual counterparts should change accordingly for a plausible experience. However, this will not be possible since the users' physical surroundings do not change during teleportation. 
To the best of our knowledge, no prior work has investigated contextualized CPS with a more flexible appearance. 
%Aerosols
Aerosols share many aspects with the requirements for CPSs, such as gradual emergence with varying degrees of opacity and dynamics~\cite{hinds2022aerosol}, and due to the intuitive transfer of real-world experience to VR~\cite{simeone2015substitutional}, users are likely to expect similar physics. We therefore developed a PCPS focusing on the timing and manner of formation as well as on the urgency-dependent appearance, to optimize both safety and presence~\cite{welch_effects_1996}.

\section{\MakeUppercase{Concept}}\label{sec:concept}
The average physical space used for VR is small: according to Barnett~\cite{barnett_steamvr_2017}, $75\%$ of play areas measure around $2\times1.5\,\mathrm{m}$, and only $0.3\%$ reach $4\times4\,\mathrm{m}$. As a CPS visualization must appear at a proximity of at least $0.4\,\mathrm{m}$ to allow a timely response, VR users may, in the worst case, be confronted with it throughout the entire VR experience. 
Hence, the appearance of boundary visualizations should be optimized to distract from the actual VE as little as possible, while still ensuring timely reactions and a sufficiently accurate estimation of the boundary's position to prevent accidental crossings. Realizing this trade-off with a sight-restriction-based visualization poses two implementational challenges. 

First, the visualization must communicate a call to stop or redirect movement without resorting to either of the two established strategies for conveying urgency: referencing to the physical environment, which reintroduces exactly the real-world anchors that threaten presence, or maximizing salience through implausible, attention-forcing stimuli, which may break the plausibility of the VE. The design space is thus narrowed to cues that are urgent yet contextualized. In our case, the state-dependent intensification of a naturally occurring phenomenon. 

Second, while sight restriction reliably impedes navigation,\newline aerosols do not inherently communicate danger: fog invites careful movement, but unlike a wall or a grid, it does not present itself as uncrossable. A particle-based boundary must therefore be designed such that users neither reach through it nor lean their head through it. For instance, through an increase of density, size, or responsiveness to suggest a volumetric barrier rather than a translucent veil. Balancing these properties is delicate: too subtle, and the visualization fails as a stopping cue. Too aggressive, and it provokes breaks in plausibility or presence, mitigating the advantage over grid-based solutions that motivated the approach.
        
\noindent Arguably, the PCPSs' visualization could simply appear at a proximity of $0.5\,\mathrm{m}$ comparable to state-of-the-art GCPS solutions. However, users might need longer to recognize the call to action, as the visualization is more contextualized than a grid-based one. Conversely, its gradual appearance might support anticipation and thus reaction time~\cite{politis2013urgency}. We therefore conducted a pre-study to investigate when boundary visualizations are first perceived and where they expect the actual boundary to be located according to the visualization. 

\section{\MakeUppercase{Pre-study}}

\begin{figure}[h!]
    \centering
    \includegraphics[width=\linewidth]{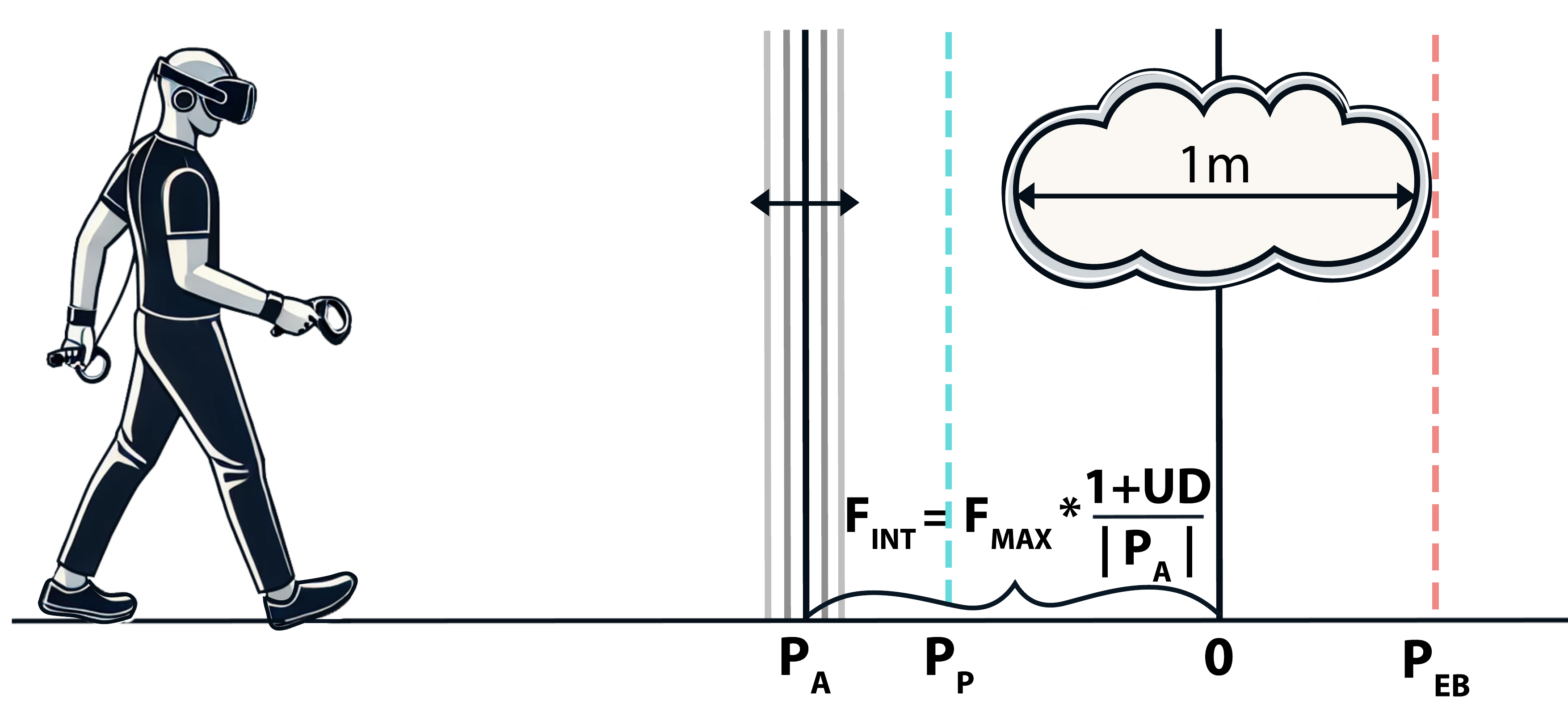}
    \caption{Settings in pre-study. Boundary visualizations appeared at a distance of 2 meters with a theoretical boundary at 3 meters distance. The visualization zone was adapted in $0.25\,\mathrm{m}$ steps according to the participants' feedback.}
    \Description{Figure 2 shows the concept of the first study. It shows a sketched person in a 2D perspective walking from left to right. The person wears a VR headset and approaches a symbolized cloud labeled with a 1-meter width. Vertical lines indicate, from left to right, the point of the cloud’s appearance (labeled with P_A), the point entered by the participants where they first perceived the cloud (labeled with P_P), the center of the cloud (labeled with 0) and the point of entered by the participants where they expected the actual boundary to be. The formula of the parameters’ intensity of the cloud is written above the range from the point of appearance to the center of the cloud, indicating the range where the parameters were increased.}
    \label{fig:boundary_pre}
\end{figure}

\noindent As a first step, we wanted to comprehend, how the necessary dynamic visual adaptations of a PCPS can be realized using standard parameters of particle systems. To retain flexibility regarding the particles' visual appearance while providing an authentic emergence~\cite{aerosols_osti_1997}, we chose to focus on \emph{density, transparency, and size}. We then conducted a pre-study to gain initial insights into which of the three parameters would best communicate an upcoming boundary or obstacle. Therefore, we investigated when boundary visualizations are perceived and when users expect the actual boundary or obstacle based on their appearance.\\

\subsection{Procedure}
We implemented a simple VR scenario and integrated a particle system resembling actual fog. The study took place in a room with a physical space of $2\times5\,\mathrm{m}$. The Meta Quest 2 with Air Link was used for the study. Participants were asked to walk straight from a starting position and press a button on the controller \emph{once when they've first perceived} the boundary visualization (onset position: $d_\mathrm{noticed}$) and once where they \emph{expect the boundary} to be according to the visualization (expected position: $d_{\mathrm{exp.}}$). Afterward, participants were further asked if they perceived the appearance of the boundary visualization as too early, right on time, or too late to notice the upcoming boundary, or if they did not perceive it as a boundary visualization at all. Furthermore, participants were allowed to comment on their experiences during the trials. The entire session was recorded.\\

\noindent The visualization started to appear at a distance between the user and the boundary of $1\,\mathrm{m}$ ($d_\mathrm{on}$). Afterward, the intensity of specific particle system parameters ($v_{attr}$) was linearly increased according to the user's distance ($\delta$) to a theoretical boundary and reached its maximum ($v_{attr}^{\mathrm{max}}$) when users reached the theoretical boundary (\autoref{fig:boundary_pre}). This process was repeated five times for the parameters density, size, and transparency. In conditions 1-3, only one of the three parameters was adapted. In the fourth condition, all three parameters were adapted simultaneously. In the fifth baseline condition, the GCPS of Meta was used~\cite{meta_set_2024}. The ordering of the conditions was randomized. For a representative experience, the theoretical boundary ($d_\mathrm{0}$) was positioned at a distance of 3 meters~\cite {barnett_steamvr_2017}. If participants stated the visualization as too early or too late, $d_\mathrm{on}$ was adjusted in $ 0.25\,\mathrm{m}$ steps accordingly, and participants repeated the process (\autoref{fig:boundary_pre}). 

\subsection{Participants}
$N=10$ people ($8$ male, $2$ female) participated in the study with a mean age of $26.1$ ($SD=1.91$). Seven said they use VR occasionally; one used it only a few times, and 2 used VR regularly for around 2-8 hours a month.

\subsection{Results}
\begin{figure}[h]
    \centering
    \includegraphics[width=\linewidth]{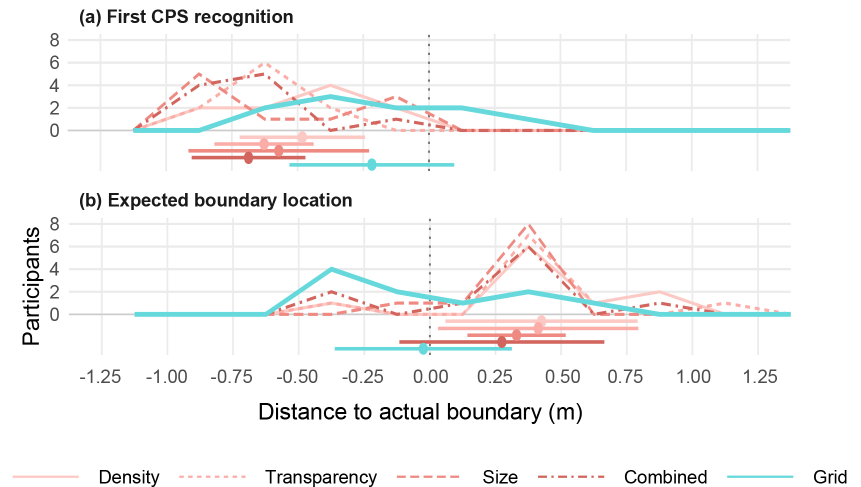}
    \caption{Distribution of the locations where participants (a) first perceived the collision prevention system's (CPS) visualization and (b) where they expected the boundary to be accordingly. The vertical dotted line shows the theoretical boundary.}
    \Description{The visualization shows two horizontal line histograms with participant count on the y-axis and distance to boundary on the x-axis. While the location of the first recognition varied greatly for all solutions, users mostly agreed about the expected location using the particle boundary.}
    \label{fig:histograms_prestudy}
    \end{figure}

The descriptives of the reported locations of perceived onset and expected boundary for all variants are summarized in~\autoref{tab:predescriptives} and illustrated in~\autoref{fig:histograms_prestudy}.\\

\subsubsection{Perceived onset}
%Most participants did not notice any differences between conditions 1-3, which was probably caused by the participants' pace, as slower participants noticed the differences. 
The reported positions of perceived onset outline the preferability of a combined solution. Users tended to notice the visualization earliest when size, density, and transparency were simultaneously increased ($Mdn=0.71\,\mathrm{m},\,IQR\,0.60-0.84,\, SD=0.04$). However, the latest reported point of first perception was $0.20$ meters, which is likely too late to prevent collisions and hence highlights the need for a more intrusive solution.

\subsubsection{Expected boundary position}
The reports of the expected boundary position further revealed that participants expected the actual boundary to be behind the theoretical boundary for all variants, with an average median of $0.40\,\mathrm{m}$ ($Mdn=0.42\,\mathrm{m},\,IQR\,0.23-0.45,\,SD=0.14$ for the combined variant). Thus, the PCPS visualization should be shifted towards the user accordingly, for exact boundary communication.

\subsubsection{Qualitative feedback}
Participants preferred the variant with increasing transparency and the combined variant, as they found them less intrusive at onset [P1-3,5-7,9]. Suggestions for design improvements mainly concerned a non-intrusive onset and more effective communication towards a critical distance to the boundary. \emph{Pre-warning:} Some wished for a light hint at a greater distance as they were nervous about colliding [P3,4]. \emph{Critical warning:} Coming closer to the boundary, many wished to have a more intrusive warning. Some suggested making the particle cloud increase until the whole FOV is covered by opaque fog [P6-8,10]; others suggested making the fog change color to a more alarming color [P5] or switching to the GCPS [P9].\\

\subsection{Discussion}
Participants preferred the combined and transparency variants, while the combined variant was perceived earlier. The pre-study further revealed that users expect the boundary to be located behind the particle cloud. Although this can be easily fixed with a shift of the visualization towards the user, this poses the question of whether users are able to accurately determine the end of the cloud. Apparently, the end of the cloud is visually occluded by the cloud itself. With a standard deviation of $14\,\mathrm{cm}$, the uncertainty about the actual boundary's location using the combined variant was rather high. While this could be attributed to interpretational differences of the boundary location relative to the visualization, it still highlights the need for clearer communication, which is in line with participants' request for more intrusive critical communication. The design should clearly communicate when users should stop to prevent collisions, regardless of how visualizations are interpreted. 
% Hence, solutions should opt for mechanisms that rely on actively preventing further approaches rather than calling for action through boundary-location communication. 
The complete cloud opaqueness and FOV coverage proposed by the participants, are mechanisms that rely on the suppression of navigational information. These solutions not only avoid interpretational effects but further do not require clear boundary location communication, which can provoke breaks in presence through reminding of the actual physical surroundings.\\

\section{\MakeUppercase{Particle-Based Collision Prevention}}    
The proposed~CPS provides a proximity-sensitive prioritization of safety and presence.

\subsection{Particle System}
The system uses a regular particle system by Unity v.23.1.18f1. The particles consist of 2D images, while the overall shape of the spawn area is a cube of $1\times1\times0.5\,\mathrm{m}$. 
To address the expectation mismatch observed in the pre-study, where participants expected the physical boundary at a median of $0.42,\mathrm{m}$ behind the visualization, the particle system's center is positioned $0.42,\mathrm{m}$ in front of the defined boundary. This shift aligns the expected boundary location with the actual one, so that users stop at the right position. All distances reported below refer to the actual boundary, not to the shifted visualization. Further configuration details are summarized in~\autoref{tab:particle-config}.

\subsection{Proximity Zones}
Following the pre-study, in which the combined variant was perceived earliest and was among the preferred variants, the PCPS manipulates three parameters of the particle system simultaneously depending on the user's proximity: density, size, and transparency. The system distinguishes three proximity zones based on the distance between the head-mounted display (HMD) and the closest boundary: a \emph{neutral zone} in which no visualization is shown ($\delta>0.8\,\mathrm{m}$), an \emph{emergence zone} spanning $d_\mathrm{on}$ to $d_\mathrm{full}$ where intensity linearly increases and a \emph{critical zone} where factor intensity is maximized ($\delta<0.4$). $\delta$ is computed as the minimum distance of the HMD to the boundary. The onset distance was set to $0.8,\mathrm{m}$, bounded from two sides by the pre-study results. Its upper bound follows from unobtrusiveness: at the $1\,\mathrm{m}$ onset used in the pre-study, several participants judged the visualization to appear too early, indicating that a sight-restricting cloud forfeits its unobtrusiveness well before that distance. Its lower bound follows from localization accuracy: participants expected the physical boundary a median of $0.42,\mathrm{m}$ behind the visualization, so the onset must leave room for this offset in addition to the distance covered between perceiving the cloud and coming to a stop.

Splitting the reaction distance into two zones relaxes the latter requirement. Because the emergence zone already signals the approaching boundary, the critical zone only has to cover the distance needed to stop once the call to act has been recognized, and can therefore be kept narrow instead of carrying the entire communication on its own. Setting both zones to $0.4,\mathrm{m}$ places the onset at $0.8,\mathrm{m}$. These values correspond to the distance covered by one average step and by half a step, respectively. Sousa et al.~\cite{sousa_safe_2019} derived comparable thresholds for their proximity-based indicators, placing the boundary of their warning zone at $0.8\,\mathrm{m}$ and of their danger zone at $0.4,\mathrm{m}$, which supports the choice independently of our pre-study.\\

\paragraph{Critical zone:} For comparable levels of safety compared to state-of-the-art GCPS, we included a zone where effective communication was prioritized over presence. Due to the linear transition from the previous stage, the distance to stop at this point should come down to users' reaction time plus the time for motoric reactions. Hence, we set the width of this zone to $0.4\,\mathrm{m}$. 

At $d_{\mathrm{full}}$, the boundary visualization covers the users' whole FOV and is completely opaque so that they can not see through anymore. The idea is to urge for redirection by disabling navigation in the boundary area unless they do so. To still allow for other navigational actions, we mapped $f$ to the delta of the user's head direction and the boundary position ($\theta$) so that $f$ linearly decreased when the user turned away from the boundary. 

\subsection{Proximity-Sensitive Visibility}
The visualization's intensity is modulated by a visibility factor $f$. Within the emergence zone, the target visibility scales linearly with proximity, so the particle cloud gradually intensifies as the user approaches. Within the critical zone, $f$ depends on the user's head orientation, where $\theta$ is the yaw angle between the viewing direction and the boundary normal. This gaze-dependent term maximizes intensity when the user faces the boundary directly, but promotes reorientation through intensity reduction when turning around. 

\noindent The factor was calculated as follows:

    \begin{equation}\label{eq:intensity}
    f(\delta,\theta) =
    \begin{cases}
        0 & \text{if } \delta \geq d_{\mathrm{on}} \\[4pt]
        1 - \dfrac{\delta - d_{\mathrm{full}}}{d_{\mathrm{on}} - d_{\mathrm{full}}} & \text{if } d_{\mathrm{full}} \leq \delta < d_{\mathrm{on}} \\[4pt]
        1 - \dfrac{\theta}{\ang{180}} & \text{if } \delta < d_{\mathrm{full}}
    \end{cases}
\end{equation}

\vspace{0.5cm}

\noindent The intensity of the different particle system attributes (size, density, and transparency) was then calculated as follows: 

 \begin{equation}
v_a = f \cdot v_a^{\mathrm{max}}, \qquad a \in \mathcal{A}_c
\label{eq:mapping}
\end{equation}

\subsection{Boundary Tracking during Locomotion} When the user moves along a boundary, the visualization glides continuously along the boundary to follow the user to realize authentic aerosol behavior and avoid implausible behavior. When the nearest side changes, it is repositioned instantaneously and the fade-in restarts from near-invisibility ($f = 0.05$) to avoid particles visibly traversing the play area. After teleportation, the tracking space is re-aligned and the visualization restored to its stored boundary offset.

\subsection{Collision Logging}
Boundary collisions were detected based on the HMD position. A collision began when the HMD position first exceeded the play area extents and ended once it returned inside. Each onset incremented the collision count, and onset and offset timestamps yielded the collision duration.

\section{\MakeUppercase{Main study}}
We then performed a within-subjects study to explore how users interact with a PCPS in comparison to the standard grid-based solution by Meta~\cite{meta_set_2024} and evaluated both with regard to safety and immersion. The GCPS has a configurable onset distance, which we set to the default of $0.5\,\mathrm{m}$. We chose a simple collection task in which participants had to collect twenty spheres one after another by walking through them. We combined natural walking and teleportation, motivated by the benefits of natural walking for presence~\cite{slater1999walking}, while accounting for the limited dimensions of the play area. While participants performed the task, we logged boundary crossings as hypothetical collisions. Physical collisions were prevented by providing a considerably larger physical space than the play area. The virtual environment consisted of an enclosed area in a virtual town (\autoref{fig:teaser}). The area was divided into five spawn zones, with two spheres present at any time. Spheres spawned in different zones, so collecting one always required moving to another zone. To prevent participants from simply teleporting from sphere to sphere and thus never approaching the boundaries of their physical environment, we let the spheres drift in random directions at walking pace after spawning. This way, participants were encouraged to walk a few steps to catch up with a sphere after teleporting. If a sphere left the enclosed area, it respawned in a different zone.

\subsection{Measures}  
    \paragraph{Position logging and number of collisions}
        During each trial, we logged the users' location and all boundary collisions, including their number, duration, and the user's position and rotation at the time of collision. A collision began when the HMD first crossed the play area boundary and ended once it had returned to within the play area. Collision duration was computed as the difference between these two timestamps.
    \paragraph{Collision Anxiety Questionnaire (CAQ)~\cite{ring_2024_caq}}
        The CAQ is a validated questionnaire to assess users' fear of colliding with real-world obstacles with the subscales \emph{general collision anxiety} (GCA), \emph{orientation} (ORI), and \emph{interpersonal collision anxiety} (IPC). It consists of 10 items answered on a 5-point Likert scale. We did not include the IPC subscale (3 items), as no interpersonal collisions were possible during the study.
    \paragraph{Slater-Usoh-Steed Questionnaire (SUSQ) \cite{usoh_using_2000,slater_depth_1994}}
        The SUSQ assesses presence in virtual environments and includes six items answered on a 7-point Likert scale with item-specific labels. The scores are calculated as the sum of all items that scored with six or seven.
    \paragraph{Safety} We additionally asked users whether they felt safe using the respective CPS with two single items, including "I was able to pay attention to the task (collecting balls)." and "I felt safe from possible collisions." Both items were rated on a 7-point Likert-Scale from "1 - strongly disagree" to "7 - strongly agree".
 
    \paragraph{Subjective Evaluation Questionnaire (SEQ)~\cite{chung_vrcat_2022}}
        In a post-study questionnaire, we asked participants about their general opinion of the PCPS. We used a subjective questionnaire based on similar studies on boundary visualizations. This included three items: "Do you think the particle cloud-based collision prevention system is useful for playing VR games safely?" (Usefulness)~\cite{chung_vrcat_2022}, "Which collision prevention system did you prefer?" (Preference)~\cite{simeone2015substitutional}, and "What needs to be improved in the particle cloud-based collision prevention system?" (Improvement)~\cite{chung_vrcat_2022}. Additionally, we asked users what they liked about the PCPS.

\subsection{Apparatus \& Procedure}
The study was conducted in a $4\times5\,\mathrm{m}$ physical tracking space. We set the virtual play area to $2\times3\,\mathrm{m}$, which lies within the range of domestic play areas reported by Barnett~\cite{barnett_steamvr_2017}. As our aim was to compare locomotion behavior between the two CPS types, we required an area in which participants could actually walk. In a smaller play area, near-constant encounters with the boundary visualization would have suppressed physical movement almost entirely. We used an Oculus Quest 2 wirelessly connected to the PC. Upon arrival, participants were asked to fill out a consent form and were introduced to the technical setup. Since most VR systems offer an initial introduction to their built-in CPS, we included a familiarization phase as well, in which participants collected three spheres, once with the PCPS and once with the GCPS. The order of the test conditions was counterbalanced. Each trial lasted up to approximately 6 minutes. After both trials, participants completed a demographic questionnaire, the SEQ, and a preference questionnaire, and were compensated for their participation. The total study time was approximately 35 minutes.

\subsection{Participants}
A total of 11 participants took part in the study. Ten participants identified as male and one as female. Ages ranged from 26 to 48 years ($M = 30.08$, $SD = 5.78$). Regarding VR experience, seven participants reported using VR occasionally (i.e., more than five times but not regularly), two had used VR only a few times, and two used VR regularly, averaging between 8 and 20 hours per month.

\subsection{Results} 
Due to the limited number of participants, we limit the evaluation of the questionnaires and number of collisions to a descriptive analysis. We further analyzed the position logs of all $11$ participants in both conditions ($\approx134,000$ samples at $\approx\SI{36}{\hertz}$). All statistics analyzing the logging data were computed based on medians and compared using Wilcoxon signed-rank tests. \\

\begin{figure}
    \centering
    \includegraphics[width=\linewidth]{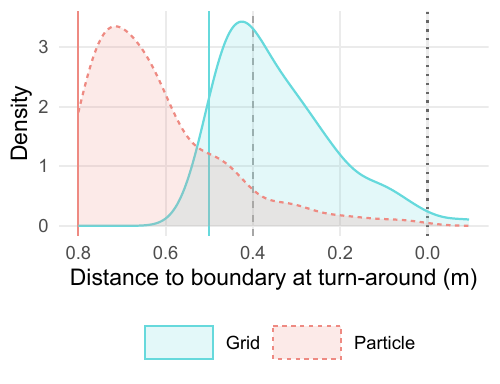}
    \caption{Distribution of turn-around distances across the approach episodes. The dotted line marks the physical boundary at 0 m; the colored vertical lines mark each system's visualization onset, the dashed line the start of the critical zone of the particle-based collision prevention system at $0.4\,\mathrm{m}$.}
    \Description{Two plots that are vertically aligned with five histograms each. The upper plot the shows the of the entered points of first perception, the second the points of expected boundary location. Each plot includes histograms for the conditions with increasing density, size, transparency, a combination of the three parameters, and the baseline, which was Meta’s Guardian system. Each histogram shows the number of participants who reported a certain position as a line showing the interpolated distribution of ratings. The x-axis represents the distance to the boundary, ranging from -1.25 to 1.25, with a dotted line indicating the boundary position at zero. The y-axis represents the number of participants. All points of first recognition of the particle-based conditions are within a 0.5 and 0.75 meters before the actual boundary on average, while it was less than 0.25 meters for the baseline CPS. The baseline on the contrary, was expected slightly before the actual boundary, while all particle-based solutions were expected at 0.25 to 0.5 meters behind the actual boundary.}
    \label{fig:turnaround}
    \end{figure}

\paragraph{Teleportation and real walking.} Locomotion strategies were similar in both conditions: participants teleported $Mdn=117$ times using the GCPS and $Mdn=100$ times using the PCPS, walking a distance of $Mdn=30.39\,\mathrm{m}$ ($IQR\,22-38$) and $Mdn=27.57\,\mathrm{m}$ ($IQR\,17-38$), respectively.

\paragraph{Visibility and Approaches.} Through the earlier onset and the limited play area, the PCPS was visible during a larger share of the session ($M = 37.0\%, SD = 11.5$) than the GCPS ($M = 12.6\%, SD = 8.4$), $V = 0$, $p = .001$, $r_{\mathrm{rb}} = 1.00$.

Approach episodes were defined as decreasing distance to the nearest boundary spanning at least $0.1\,\mathrm{m}$. Since a participant cannot react to a visualization that has not appeared, only episodes crossing the respective onset distance were analyzed, yielding $253$ approach episodes (GCPS: $85,~Mdn = 4$,\,IQR\,$2-12$; PCPS: $168,~Mdn = 17,\,IQR\,9.5-21$).

\paragraph{Turn-around distance.} In most of the approaches, participants stopped or redirected after visualization onset (PCPS: $M = 0.65\,\mathrm{m},\newline SD = 0.07$; GCPS: $M = 0.36\,\mathrm{m},~SD = 0.08$) $V = 0$, $p = .001$, $r_{\mathrm{rb}} = 1.00$ (see~\autoref{fig:turnaround}). As the turn-around distance is bounded by the respective onset distance, this difference primarily reflects the earlier onset of the PCPS rather than a difference between the visualizations themselves.

\paragraph{Reaction distance.} Aligning both conditions to their onset, participants continued approaching the visualization by an equivalent distance before stopping (GCPS: $M = 0.14\,\mathrm{m},~SD = 0.08$; PCPS: $M = 0.15\,\mathrm{m},~SD = 0.07$; $V = 25$, $p = .520$, $r_{\mathrm{rb}} = 0.24$, $95\%\,\mathrm{CI}\,[-0.48, 0.82]$). The approach pace itself did not differ between conditions (GCPS: $M = 16.3 m/min,~SD = 4.6$; PCPS: $M = 15.6 m/min,~SD = 3.1$; $V = 20$, $p = .278$, $r_{\mathrm{rb}} = -0.39$, $95\%\,\mathrm{CI}\,[-0.91,\newline0.27]$), and showed no gradual deceleration after onset in either condition, indicating that participants maintained their pace and then stopped rather than slowing down progressively. %(see~\autoref{fig:penetration}).
 
\paragraph{Collisions.} Using the GCPS, participants collided around $0.18$ times on average with the theoretical boundaries of the play area ($CI=[-0.05,0.41],~SD=0.39$). When using the PCPS, participants collided with the theoretical boundary around $0.36$ times (${CI=[-0.09,0.82]},~SD=0.77$). We additionally asked users whether they felt safe using the respective CPS. The perceived safety using the GCPS was rated with $M=4.18$ ($SD=2.27,~CI=[2.84,5.52]$) and the PCPS with $M=5.09$ ($SD=2.07,~CI=[3.87,6.31]$). A Bayesian equivalence paired-samples t-test (equivalence region $\delta \in [-0.47, 0.47]$, corresponding to $\pm 1$ scale point; Cauchy prior $r = 0.707$) yielded $\mathrm{BF}_{\in\notin} = 6.40$, indicating moderate evidence that perceived safety was equivalent between the GCPS and the PCPS (posterior median $\delta = 0.24$, $95\%\,\mathrm{CI}\,[-0.30, 0.79]$).
Participants further reported that they were able to focus on the task using both solutions (GCPS: $M=6.36,~SD=0.81,~CI=[5.88,6.84]$, PCPS: $M=6.36,~SD=1.21,~CI=[5.65,7.08]$). 

\paragraph{Presence}

\begin{figure*}
    \centering
    \includegraphics[width=\linewidth]{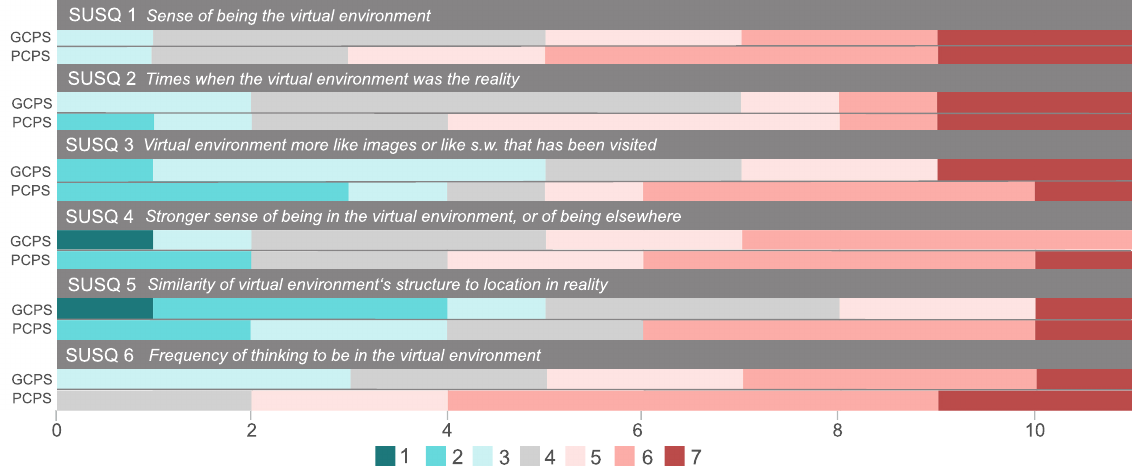}
    \caption{Ratings of the Slater-Usoh-Steed questionnaire per item split by collision prevention system (grid-based collision prevention system = GCPS, particle-based collision prevention system = PCPS.}
    \Description{A stacked bar chart of the ratings of the individual items of the Slater-Usoh-Steed Questionnaire. The items are arranged on the y-axis from top to bottom, grouped by condition, each having a short description of the item content as a header line. The x-axis is the count of answers ranging from zero to eleven.}
    \label{fig:presence_likert}
    \end{figure*}

Participants reported higher levels of presence when using the PCPS ($Mdn=3,\,IQR\,0.5-5$) compared to the GCPS ($Mdn = 1,\,IQR\,0-3$). Participants' ratings are illustrated in~\autoref{fig:presence_likert}.

\paragraph{Collision Anxiety}
\begin{figure}
    \centering
    \includegraphics[width=\linewidth]{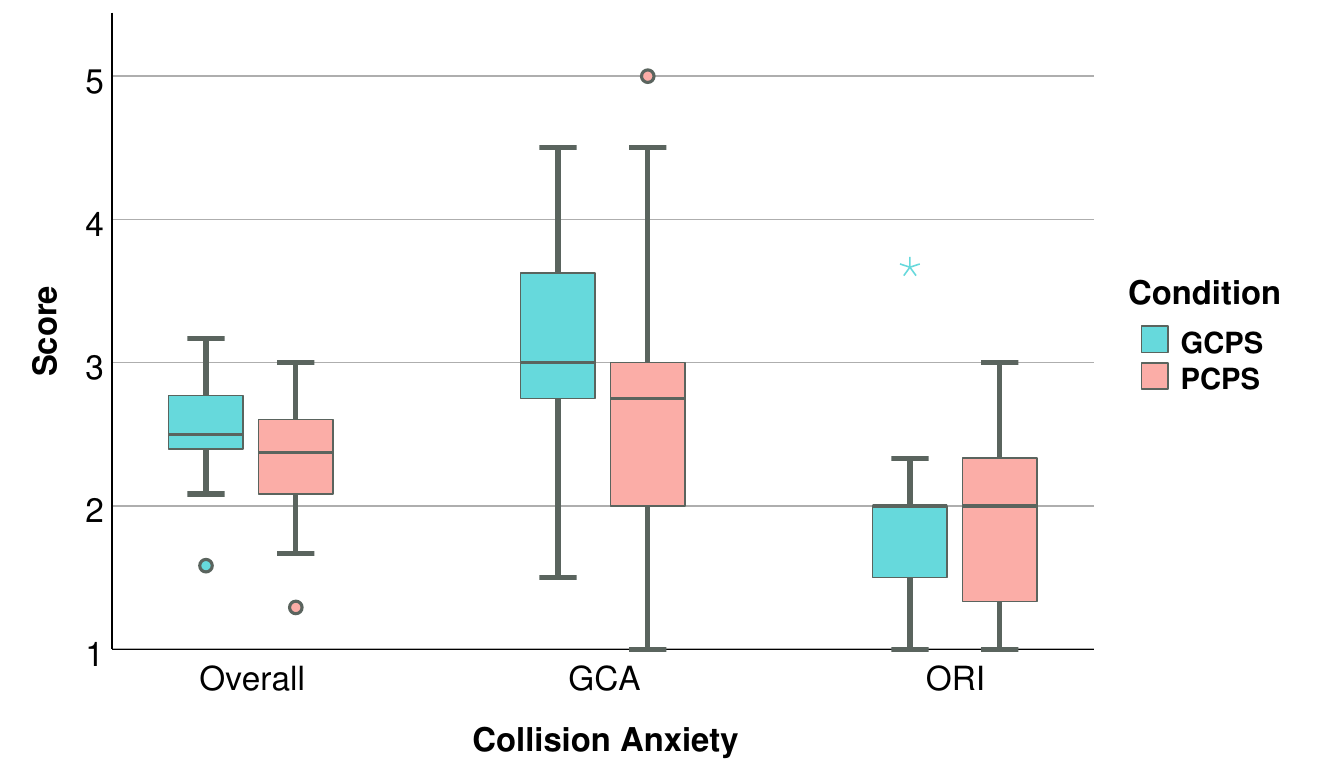}
    \caption{Boxplots of the scores of the Collision Anxiety Questionnaire grouped by CPS type. Left: the overall collision anxiety score. In the middle: the results of the general collision anxiety subscale. Right: the results of the orientation subscale.}
    \Description{Boxplots of the scores of the Collision Anxiety Questionnaire. They are grouped from left to right by overall score, the subscale “General Collision Anxiety”, and the subscale “Orientation”. Each group includes a boxplot for the “GCPS” condition and the “PCPS” condition. The y-axis represents the scores from 1 to 5. Whiskers indicate the 90\% confidence interval. Dots indicate outliers. A line represents the average score. The overall and general collision anxiety scores are slightly higher in the grid-based condition, while orientation shows no difference. The deviation of general collision anxiety is notably higher than of the overall scores.}
    \label{fig:boxplot_ca}
    \end{figure}
    
Reported collision anxiety was similar in both conditions (GCPS: $Mdn=3.0,\,IQR\,2.96-3.62$; PCPS: $Mdn=2.71,\newline IQR\,2.0-3.0$), as were the GCA and ORI subscales (see~\autoref{fig:boxplot_ca}).

\paragraph{Preference}
The overall feedback on the PCPS was positive. Participants were also asked about the utility of a PCPS for VR gaming safety. Five participants strongly agreed, while five participants agreed or mildly agreed that the PCPS was useful. One mildly disagreed with its usefulness. Seven of the eleven participants stated that they preferred the PCPS. Six participants reported that the GCPS more often reminded them of their real-world surroundings.

\section{\MakeUppercase{Limitations}}
Our study has some limitations. As we wanted to focus our research on how users interact with PCPS compared to grid-based solutions, we used a simple rectangular play area and did not consider obstacles other than the boundaries themselves. Although we used common dimensions for our play area, this limits external validity.
Additionally, our sample was small ($N = 11$) and predominantly male, limiting the generalizability of our findings. Future studies should recruit larger, more diverse samples and extend the comparison to non-rectangular play areas and play areas containing interior obstacles. Another limitation is that we operationalized boundary collisions via the position of the HMD only. Crossings of the controllers with the boundary were therefore not captured. However, because participants had to collect spheres with the HMD rather than with the controllers, the task design prevented participants from simply reaching through the boundary to collect spheres. Furthermore, as this operationalization was applied identically across all conditions, it does not affect the validity of the comparison between the PCPS and GCPS.
Occasional performance drops in wireless streaming may have affected the perceived smoothness of the particle visualization. The task could further be completed solely through micro-teleportation; we intentionally permitted this to capture realistic locomotion behavior. Logging showed two participants primarily teleported, yet both encountered the CPS at least once, and mixed locomotion preferences mirror real-world usage~\cite{tseng_understanding_2024}.

\section{\MakeUppercase{Discussion}}
 
We conducted two studies on our three research questions. In an initial pre-study, we investigated how users perceive and interact with particle clouds to determine how they should be designed to most effectively and precisely communicate upcoming boundaries of physical space \textbf{(RQ1)}. The second study aimed to compare our design to a grid CPS towards locomotion behavior when approaching boundaries and collisions \textbf{(RQ2)}. Third, we measured presence to compare both solutions in terms of immersiveness \textbf{(RQ3)}.

\paragraph{Sight restriction functions as a stopping cue.}
Although crossing may be challenged through sight restriction, fog-like clouds are typically traversable without harm. In gaming, people are partly even rewarded for doing so by unlocking new map regions or getting rare items. P1, for example, explained that, as used in games, one can reach through foggy areas without harm and that it might be rewarded with something interesting within or behind the fog. Accordingly, we explored whether it is even possible to evoke stops or redirections working with such boundary visualizations. Our main study illustrated that the use of standard particle systems, although translucent and non-solid, may evoke similar CPS interaction patterns to grid-based solutions: participants interrupted their approach after a comparable distance in both conditions, and their pace when approaching the boundary did not differ. This was the case even though participants in the pre-study expressed uncertainty about the exact boundary location using the PCPS. 

Our design counteracts the traversability of fog by achieving complete opacity at close range, thereby impeding navigation~\cite{davis_investigating_2023}. Introducing clearer design features, such as a color shift to red, might further enhance this effect but could also risk decreasing the system's plausibility and immersiveness.

\paragraph{The emergence phase was sufficient.}
Under the PCPS, participants turned around at around $0.15\,\mathrm{m}$ past the onset and thus within the outer half of the $0.4\,\mathrm{m}$ emergence zone. According to our visibility model (Eq.~\ref{eq:intensity}), the particle cloud had reached only about $40\%$ of its maximum target intensity at this point, and less in practice, as the fade-in easing delays the realized intensity behind its target. This addresses the first challenge outlined in~\autoref{sec:concept}: communicating a call to act did not require a salient stimulus. A faint, semi-transparent cloud that had barely begun to materialize was sufficient to redirect movement, suggesting that the mere appearance carries the signal rather than the intensity of what appears. The absence of gradual deceleration when approaching a boundary points the same way: participants responded to the onset itself instead of tracking the visualization's growing intensity.

\paragraph{Safety and intentional boundary crossings.}
Participants reaction distances and turn-around behavior using the PCPS evidenced comparable safety levels to those of state-of-the-art GCPS, namely Meta's Guardian. The designed PCPS was capable of provoking the desired movement redirection. Boundary crossings were rare under both systems: two participants crossed the boundary while using the GCPS and two while using the PCPS. Given these numbers, the collision counts cannot support a comparative claim, and our safety argument rests on the approach behavior reported above. Nevertheless, participants reported slightly lower collision anxiety using the PCPS, suggesting that some users even perceived it as safer. We attribute this effect to the increased depth of the PCPS visualization and the lack of information about the boundaries' exact locations. According to participants' qualitative feedback, they perceived the available physical space as smaller (P4, P6), which presumably led them to expect more space until they reached a boundary. Interestingly, they nevertheless exhibited similar movement patterns with an average movement radius of $1.19\,\mathrm{m}$, while it was $1.21\,\mathrm{m}$ using the GCPS (see also~\autoref{fig:heatmaps}).

Despite this misperception of a greater distance to the boundaries, only two participants collided with the theoretical boundaries while using the PCPS. We assume these boundary crossings were performed intentionally due to a need for more precise information on the boundaries' locations: \textit{``I missed a similar visualization [showing] where or which body part left the boundary as it is the case for the grid-based approach''} (P5).
As participants generally reported that they felt able to orient themselves within the physical environment (see~\autoref{fig:boxplot_ca}), we hypothesize that these collisions could represent intentional boundary crossings. Prior research by Tseng et al.~\cite{tseng_understanding_2024} highlights the phenomenon of users deliberately breaching safety boundaries in VR when they have a mental model of the physical environment that suggests no real danger. In line with this, our participants who collided more frequently with the PCPS were those with greater prior VR experience, suggesting a level of familiarity with CPS cues. Moreover, we placed the spheres to collect at eye level, and participants probably remembered that there were no obstacles within the physical space at that height. This may have reinforced their decision to ignore the PCPS's warnings.

\paragraph{Ambiguity, real-world associations, and presence.}
The GCPS communicates spatial boundaries with high precision, showing users the exact layout of their configured play area. While effective for safety and spatial awareness, this explicit representation also evokes the user's mental model of the real world, as the shape of a configured boundary typically reflects features of the physical space. Greater detail about the exact shape and position of boundaries thus makes users aware of real-world information, which in turn may reduce presence. Six participants reported that the grid reminded them of their physical room. The PCPS, by contrast, communicates direction and urgency without representing precise boundaries, and its ambiguous visual structure helps avoid real-world associations.

This is reflected in our results: although the PCPS was visible three times as often as the GCPS, it still evoked higher levels of presence, and participants reported feeling safer. Participants appreciated its immersive qualities, describing it as \textit{``smooth and natural''} (P3), \textit{``more pleasant''} (P7), and \textit{``more `realistic' and [\ldots] less `cyberspace'\,''} (P5). Participants perceived the fog as less intrusive and more thematically appropriate for virtual settings. Although the pre-study showed that the visualization's looser correspondence to the boundary shape reduced localization accuracy, the effects on safety were largely positive: participants redirected their navigation as soon as they noticed the PCPS, and due to its earlier onset, they stayed farther from the boundary than with the GCPS. As a result, the PCPS mostly remained unobtrusive. These characteristics suggest that the PCPS supports presence without compromising safety, which aligns with the findings of Valentini et al.~\cite{valentini_improving_2020}.

\paragraph{Locomotion behavior and area utilization.}
A potential trade-off of this design is that users may feel less confident making full use of the available movement space. Participants reported a reduced perceived movement radius with the PCPS, and the tracking data show shorter walking distances in this condition. In typical domestic play areas, which are often as small as $3\times2\,\mathrm{m}$~\cite{barnett_steamvr_2017}, this perceived limitation might affect the overall experience. Notably, four participants preferred the GCPS, possibly due to a desire for more precise spatial cues. Some of them also reported the highest collision anxiety in the sample, with scores between $3.5$ and $4.5$. Others described themselves as experienced VR users who wanted to know the extent of their physical space in order to use it as efficiently as possible. Our results thus suggest two user groups: the first prioritizes an immersive experience and is willing to compromise freedom of movement in exchange for a more implicit awareness of the physical environment. The second prefers explicit spatial awareness over presence, due to either increased collision anxiety or the desire to exploit the available physical space when navigating in VR. Future variants of the PCPS might therefore offer options for users who prioritize spatial awareness over presence, such as adaptive proximity-based boundary shaping, in which the boundary becomes more sharply defined as the user approaches it.

\subsection{Design Implications}
Across both studies, our findings translate into four implications for the design of immersive CPSs:

\begin{itemize}
    \item \textbf{Onset distance is the primary lever, not visual salience.} (Main study) Both systems resulted in similar turn-around distances after onset, although the grid-based approach relies more on visual distinctiveness and increased saliency. Increasing the salience of a visualization to stop users earlier may therefore be less effective than simply showing a subtler one sooner, and may further risk the presence benefit that motivated the subtler design in the first place.

    \item \textbf{Unobtrusive visualizations allow earlier onsets.} (Pre-study and main study) Because a grid relies on visual distinctiveness from the VE whenever it is visible, its onset must be kept late to limit its presence cost. Our sight-restriction-based approach was not subject to this constraint: the PCPS was visible three times as often as the GCPS and still yielded higher presence. Designers can therefore trade exposure time for safety margin, a margin that is key given the reduced localization accuracy observed in our pre-study.

    \item \textbf{The emergence phase should carry the communication.} (Main study) Participants typically stopped while the PCPS visualization was still faint, well before it reached full intensity. The gradual, semi-transparent phase was thus sufficient to trigger redirection, and the maximally salient state functioned as a fallback rather than as the primary signal. While we used a $0.4\,\mathrm{m}$ wide critical zone to ensure users' safety, the effectiveness of the emergence phase suggests that this zone can be narrowed and the emergence phase widened accordingly, oriented at the observed turn-around distances.

    \item \textbf{Spatial precision should be optional, not absent.} (Main study) The users who preferred the GCPS did so for its precise spatial cues, and the few observed boundary crossings were plausibly intentional attempts to obtain exactly that information. Immersive CPSs should therefore offer an on-demand or adaptive way to reveal the boundary's precise location, rather than withholding it entirely.
\end{itemize}

\section{Future Work}
Our results emphasize the need for techniques that communicate boundaries precisely yet immersively, without reminding users of their physical surroundings. We showed that effective collision prevention is possible even when associations with real-world information are minimized. However, participants requested clearer and more precise communication to navigate VR confidently using real walking. Our results further showed the benefit of reusing known dynamic phenomena from reality to communicate boundaries, since such visualizations must be highly flexible in their state-dependent appearance while remaining plausible to avoid breaks in presence. Consequently, we encourage future research on how the characteristics of known dynamic phenomena can be utilized to communicate the need for action more precisely while minimizing real-world anchors.

Some participants asked for a pre-warning in the form of an unobtrusive early boundary visualization, reporting that they slowed down out of concern about missing a warning. As we targeted a design solution for realistic domestic play areas~\cite{barnett_steamvr_2017}, integrating a pre-warning would have resulted in a permanently visible boundary, so we deliberately omitted it. Nevertheless, users with high collision anxiety may prefer a permanently visible boundary to navigational uncertainty.

For comparability, we limited our design solution to visual feedback, as most existing solutions are solely visual~\cite{meta_set_2024,barnett_steamvr_2017,scavarelli2017VRcollide}. However, several studies have shown the benefits of other sensory modalities~\cite{Li2013UrgencyTactile,bridget2012multimodal,valkov2019vibrotactile}: vibrotactile and auditory feedback reduced intentional crossings of virtual obstacles~\cite{boldt1028intrusive}, and multimodal feedback can communicate urgency and reduce reaction times~\cite{politis2013urgency}. We therefore encourage research on multimodal collision prevention, particularly on whether non-visual channels can deliver the precision our participants requested without adding visual real-world anchors.

\section{Conclusion}
State-of-the-art CPS are predominantly grid-based, which communicates upcoming obstacles efficiently but likely breaks presence. Drawing on sight restriction as a mechanism known from reality to provoke cautious behavior, we designed a particle-based boundary visualization inspired by aerosols and compared this PCPS with a standard GCPS.
Although fog-like clouds are translucent and traversable without harm, participants stopped after a comparable distance once the visualization appeared, typically while the particle cloud was still faint. The mere appearance of the visualization thus seems to carry the signal, rather than its intensity. Perceived safety was equivalent in both conditions, yet most participants preferred the PCPS and reported higher presence with it, even though it was visible three times as long as the GCPS.
Contrary to the common assumption that a CPS must break presence to keep users within their physical space, our findings indicate that safety and presence are not mutually exclusive. Since limited domestic play areas make frequent encounters with boundary visualizations unavoidable, we strongly encourage further exploration of immersive CPS designs, with dynamic phenomena from reality being particularly promising. To support this, we release our CPS as an open-source Unity solution that promotes the creation and testing of new particle-based boundary visualizations.

\section*{Generative AI Disclosure}
During the preparation of this work, the authors used generative AI tools
solely to iterate on phrasing, check grammar, and assist with formatting.
All ideas, research, analysis, results, and conclusions presented in this
paper are the authors' own original work. The authors reviewed and
edited all AI-assisted text and take full responsibility for the content
of this publication.

\bibliographystyle{ACM-Reference-Format}
\bibliography{Bibliography}

\appendix

\begin{table*}[]
    \centering
    \begin{tabular}{p{3cm} c c c c c c} \hline
         & \multicolumn{3}{c}{First perceived} & \multicolumn{3}{c}{Expected boundary location}\\ \cmidrule(lr){2-4}\cmidrule(lr){5-7}
         & Median (IQR) & Min & Max & Median (IQR) & Min & Max \\ \hline
        Density & $-0.49~(-0.53, -0.37)$ & $-0.88$ & $-0.14$ & $0.40~(0.34, 0.59)$ & $-0.39$ & $0.99$\\
        Transparency & $-0.60~(-0.73, -0.55)$ & $-1.00$ & $-0.33$ & $0.43~(0.37, 0.49)$ & $-0.44$ & $1.15$\\
        Size & $-0.65~(-0.85, -0.27)$ & $-0.95$ & $-0.05$ & $0.37~(0.34, 0.45)$ & $-0.05$ & $0.50$\\
        Combined & $-0.71~(-0.84, -0.60)$ & $-0.96$ & $-0.20$ & $0.42~(0.23, 0.45)$ & $-0.41$ & $0.83$\\
        Grid & $-0.20~(-0.48, 0.02)$ & $-0.62$ & $0.32$ & $-0.06~(-0.27,0.26)$ & $-0.46$ & $0.53$ \\ \hline       
    \end{tabular}
    \caption{Results of the pre-study showing minimum, maximum and median with interquartile range (IQR) of the distances to the actual boundary (a) where users first perceived the visualization of the collision prevention system and (b) where they expected the boundary to be.  (in meters to boundary)}
    \label{tab:predescriptives}
\end{table*}

\begin{table*}[]
  \caption{Configuration of the Unity particle system used for the particle-based collision prevention system (PCPS) visualization. The final PCPS implementation is accessible under \href{https://github.com/annalisaDegenhard/ParticleBasedCollisionPrevention.git}{Github}. Properties marked with $f$ were modulated at runtime by the intensity factor (\autoref{eq:intensity}); all listed values refer to the maximum intensity $v_a^{\mathrm{max}}$. All remaining properties followed the Unity default configuration (Unity~23.1.18f1).}
  \label{tab:particle-config}
  \begin{tabular}{@{}lll@{}}
    \toprule
    Module & Property & Value \\
    \midrule
    Main & Duration & 10\,s \\
         & Looping & enabled \\
         & Start lifetime & 5\,s \\
         & Start speed & 0/0.2\,m/s \\
         & 3D Start size & $0.8\times0.8\times0.4$\,m \\
         & Start color & (255,255,255) \\
         & Gravity modifier & 0 \\
         & Simulation space & Local \\
         & Max particles & 1000\\
         & Culling Mode & Automatic\\
    \addlinespace
    Emission & Rate over time ($f$) & $100/200 \cdot f$ particles/s \\
    \addlinespace
    Shape & Shape & Box \\
          & Scale ($f$) & $1 + 0.5f \times 1 + 0.5f \times 0.5$\,m \\
          & Randomize Direction & 1\\
    \addlinespace
    Color over lifetime & Color & White \\
          & Alpha gradient ($f$) & $0 \rightarrow 0.75 \cdot f \rightarrow 0$ \\
          & & (peak at 50\,\% lifetime) \\
    \addlinespace
    Collision & Type & Planes\\
              & Planes & Bottom, front, and back plane\\
              & Bounce & 2 \\
              & Lifetime Loss & 1\\
    \addlinespace
    Renderer & Render mode & Vertical Billboard \\
             & Material & \textit{See Github Repository} \\
    \bottomrule
  \end{tabular}
\end{table*}

\begin{figure*}
    \centering
    \includegraphics[width=\linewidth]{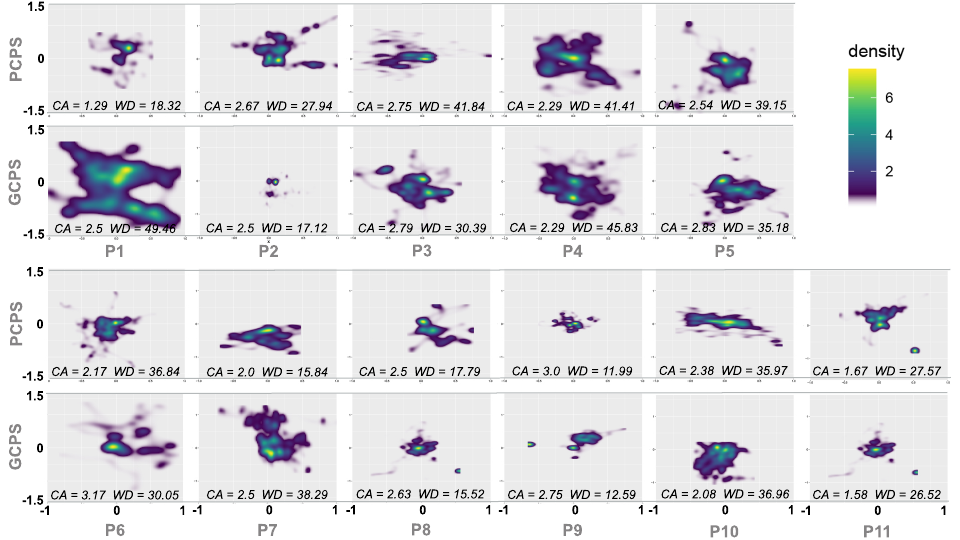}
    \caption{Heatmaps of all participants per condition. Each map shows the positions of a participant within the play area ($3\times2m$). The participants' reported collision anxiety level (CA) and the distance walked by the participants (WD) during the respective condition are displayed below.}
    \Description{For each of the eleven participants one heatmap represents the GCPS and one the PCPS condition. The dimensions of each heatmap represent the dimensions of the used play area of 2x3 meters. Each heatmap shows where participants were located within this area during the condition. A legend explains the different colors that indicate the data density for the specific location. For each heatmap, the participant’s collision anxiety score and the walked distance is displayed below.}
    \label{fig:heatmaps}
    \end{figure*}

\end{document}